\documentclass[]{pasj01}

\usepackage{graphicx}
\usepackage{times}
\usepackage{lscape}
\usepackage{color}

\begin{document}

\Received{}
\Accepted{}

\title{Application of the Ghosh \& Lamb Relation to the Spin-up/down Behavior in the X-ray Binary Pulsar 4U 1626--67}

\author{Toshihiro \textsc{Takagi}\altaffilmark{1,2,*}}
\email{takagi@crab.riken.jp}

\author{Tatehiro \textsc{Mihara}\altaffilmark{2}}
\author{Mutsumi \textsc{Sugizaki}\altaffilmark{2}}
\author{Kazuo \textsc{Makishima}\altaffilmark{2}}
\author{Mikio \textsc{Morii}\altaffilmark{3}}

\altaffiltext{1}{Department of Physics, Nihon University, 1-8-14 Kandasurugadai, Chiyoda-ku, Tokyo 101-8308, Japan}
\altaffiltext{2}{MAXI team, RIKEN, 2-1 Hirosawa, Wako, Saitama 351-0198, Japan}
\altaffiltext{3}{Research Center for Statistical Machine Learning, Institute of Statistical Mathematics, 10-3 Midori-cho, Tachikawa, Tokyo 190-8562, Japan}


\maketitle

\begin{abstract}
We analyzed continuous MAXI/GSC data of the X-ray binary pulsar 4U 1626--67 from 2009 October to 2013 September,
 and determined the pulse period and the pulse-period derivative for every 60-d interval by the epoch folding method.
The obtained periods are consistent with those provided by the Fermi/GBM pulsar project.
In all the 60-d intervals, the pulsar was observed to spin up, with the spin-up rate positively correlated with the 2--20 keV flux.
We applied the accretion torque model proposed by Ghosh \& Lamb (1979, ApJ, 234, 296) to the MAXI/GSC data,
 as well as the past data including both spin-up and spin-down phases.
The Ghosh \& Lamb relation was confirmed to successfully explain the observed relation between the spin-up/down rate and the flux.
By comparing the model-predicted luminosity with the observed flux,
 the source distance was constrained as 5--13 kpc, which is consistent with that by Chakrabarty (1998, ApJ, 492, 342).
Conversely, if the source distance is assumed, the data can constrain the mass and radius of the neutron star,
 because the Ghosh \& Lamb model depends on these parameters.
We attempted this idea, and found that an assumed distance of, e.g., 10 kpc gives a mass in the range of 1.81--1.90 solar mass, and a radius of 11.4--11.5 km,
 although these results are still subject to considerable systematic uncertainties other than that in the distance.
\end{abstract}

\section{Introduction}
An X-ray binary pulsar is a system consisting of a magnetized neutron star and a stellar companion.
The X-ray emission is powered by gravitational energy of accreting matter.
In a system with a low-mass companion star, the accretion flow is considered to take place via Roche-Lobe overflow,
 and to form an accretion disk in the vicinity of the pulsar.
The angular momentum of the accreting gas is transferred to the pulsar at the inner edge of the accretion disk,
 and accelerates the pulsar rotation until it finally reaches an equilibrium determined by the accretion rate and the magnetic moment of the pulsar.
So far, numbers of theoretical studies have been performed on the interaction between the pulsar magnetosphere and the accretion flows.
Along this scenario, \citet{1977Natur.266..683R}, \citet{1979ApJ...234..296G} (GL79 hereafter) and \citet{1995MNRAS.275..244L} (LRB95 hereafter)
 proposed their accretion models and presented equations describing the pulse-period derivative $\dot{P}$
 as a function of the luminosity $L$, the pulse period $P$, the mass $M$, the radius $R$ and the surface magnetic field ${B}_{ \rm{c} }$ of the neutron star.
These models have been examined against observational data
 (e.g. \cite{1984ARA&A..22..537J}; \cite{1996A&AS..120C.209F}; \cite{1996A&A...312..872R}; \cite{1997ApJS..113..367B}; \cite{2009A&A...506.1261K}; \cite{2015PASJ..tmp..210S}).
The results show that the observed $\dot{P}$--$L$ relations are grossly consistent with the model predictions.
However, the validity of the models has not yet been fully confirmed, because it requires long-term monitoring of some suitable objects with known ${B}_{ \rm{c} }$,
 covering significant pulse-period and luminosity changes with a sufficient sampling rate.

When these models describing the accretion torque are better calibrated, the observed $\dot{P}$--$L$ relations can give us observational constraints on $M$ and $R$,
 which are very important because they are directly connected to the equation of state (EOS) of nuclear matter.
So far, various observational attempts to measure $M$ and $R$ have been carried out, and generally yielded $M \simeq 1.4 {M}_{\odot}$ and $R \simeq 12$ km
 (e.g. reviews by \cite{2010AdSpR..45..949B} and \cite{2013RPPh...76a6901O}).
However, the measurements are not yet accurate enough to constrain the EOS.
Furthermore, the values of $M$ (mainly from X-ray binary pulsars) and $R$ (mainly from weakly-magnetized neutron stars)
 have been derived from different populations of neutron stars.
Therefore, further studies of the $\dot{P}$--$L$ relations are expected to be valuable.

4U 1626--67 is a low-mass X-ray binary pulsar first detected with the Uhuru satellite \citep{1972ApJ...178..281G},
 and its 7.6-s coherent pulsation was discovered by \citet{1977ApJ...217L..29R}.
Because no period modulation due to orbital motion has been detected beyond an upper limit of ${a}_{ \rm{x} } \sin i \le 13$ lt-ms
 (${a}_{ \rm{x} }$ is the orbital semi-major axis of the pulsar and $i$ is the orbital inclination),
 the mass of the companion star is estimated to be very low ($\sim 0.03 - 0.09 {M}_{\odot}$ for ${11}^{\circ} \leq i \leq {36}^{\circ}$; \cite{1988ApJ...327..732L}).
It is hence classified as an ultra compact X-ray binary \citep{2012A&A...543A.121V}.
The BeppoSAX observation revealed a cyclotron resonance scattering feature at $\sim 37 ~ { \rm{keV} }$,
 indicating the surface magnetic field of ${B}_{ \rm{c} } = 3.2 \times {10}^{12} ~ (1 + {z}_{ \rm{g} }) ~ { \rm{G} }$,
 where ${z}_{ \rm{g} }$ is the gravitational redshift,
\begin{equation}
{z}_{ \rm{g} } = \left(1 - \frac{2 G M}{R {c}^{2}} \right)^{ - \frac{1}{2} } - 1,
\label{equ:z_g}
\end{equation}
represented by the gravitational constant $G$ and the velocity of light $c$ \citep{1998ApJ...500L.163O}.
The feature was confirmed by the Suzaku observation \citep{2012ApJ...751...35I}.
The source distance was estimated to be 5--13 kpc
 from the optical flux by assuming that the effective X-ray albedo of the accretion disk is high ($\gtrsim$0.9) \citep{1998ApJ...492..342C}.

\begin{figure}
\begin{center}
\includegraphics[width=70mm,angle=-90]{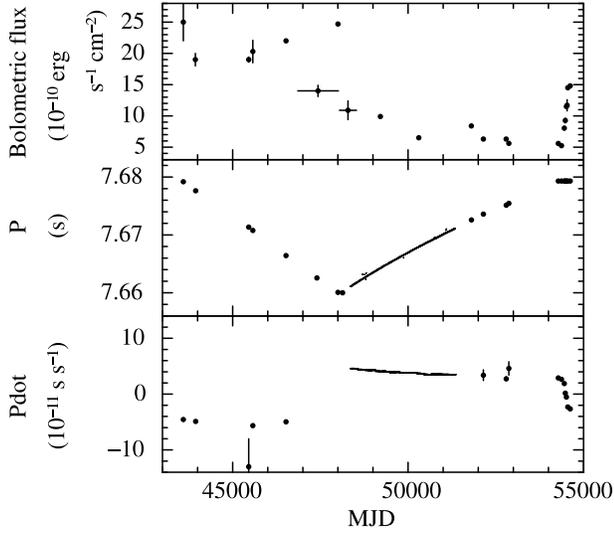}
\end{center}
\caption{Bolometric flux (top), period (middle) and period derivative (bottom) of 4U 1626--67 obtained by past X-ray observations from 1978 to 2008.
In the top panel, observed X-ray intensities are converted to the model flux in the 0.5--100 keV band assuming the typical spectral model given by \citet{2012A&A...546A..40C}.
Filled circles and solid lines represent the past observations in table \ref{tab:past_fx_p_pdot} and BATSE\footnotemark observations, respectively.}
\label{fig:all_fx_p_pdot}
\end{figure}

Since the discovery of the 7.6-s pulsation in 1977,
 the period of 4U 1626--67 has been repeatedly measured with various X-ray satellites
 (e.g. references in \cite{1997ApJ...474..414C}; \cite{2010ApJ...708.1500C}).
Table \ref{tab:past_fx_p_pdot} summarizes the X-ray fluxes, periods and period derivatives observed from 1978 to 2008,
 and figure \ref{fig:all_fx_p_pdot} visualizes long-term behavior of these quantities.
It clearly shows that the source made transitions twice between the spin-up and the spin-down phases,
 at MJD $\sim$ 48000 (1990 June) and MJD $\sim$ 54000 (2008 February), separated by $\sim$ 18 years.
In each phase $\dot{P}$ were almost constant, and its absolute values were very similar as $|\dot{P}| = 2 \sim 5 \times {10}^{-11} ~ \rm{ s ~ {s}^{-1} }$.
These period-change behavior suggests that 4U 1626--67 is close to an equilibrium state
 in which the net torque transfer from the accreting matter to the pulsar is approximately zero.
At the last transition in 2008 when the source turned from the spin-down into the spin-up phase, the flux increased by a factor of $\sim$2.5 \citep{2010ApJ...708.1500C}.
These properties, together with the accurate knowledge of ${B}_{ \rm{c} }$, make this object ideal for our study.

Monitor of All-sky X-ray Image (MAXI; \cite{2009PASJ...61..999M}) is an X-ray all-sky monitor on the International Space Station.
Since the in-orbit operation started in 2009 August, its main instrument, the GSC (Gas Slit Camera; \cite{2011PASJ...63S.623M}; \cite{2011PASJ...63S.635S}),
 has been scanning the whole sky every 92 min in the 2--20 keV band.
The GSC field of view typically scans a celestial point source for about 60 s in each transit, which is long enough to study the 7.6-s pulsation from 4U 1626--67.
Thus, the MAXI/GSC data are useful to study the long-term variation of the flux, $P$, and $\dot{P}$.

In this paper, we analyze the MAXI/GSC data of 4U 1626--67 from MJD 55110 (2009 October 6) to MJD 56550 (2013 September 15) and determine the flux, $P$, and $\dot{P}$.
We then apply the spin-up/down models proposed by GL79 and LRB95 to the previous and the MAXI/GSC data,
 to examine whether either model can explain the observed behavior of 4U 1626--67,
 and to evaluate how these data can constrain the source distance, as well as the mass and radius of the neutron star.

\begin{table*}
\caption{X-ray flux, period and period derivative obtained in past observations of 4U 1626--67.}
\label{tab:past_fx_p_pdot}
\begin{center}
\begin{tabular}{lcccccllcc}
\hline
\multicolumn{1}{l}{Observation} & \multicolumn{4}{c}{Flux} & & \multicolumn{3}{c}{Pulsation} & \\
\cline{2-5}\cline{7-9}
\multicolumn{1}{c}{Date} & \multicolumn{1}{c}{Period} & Band & Obs. & Bolometric\footnotemark[$*$]
 & & \multicolumn{1}{c}{Epoch} & \multicolumn{1}{c}{$P$\footnotemark[$\dagger$]} & $\dot{P}$ & Ref.\footnotemark[$\ddagger$] \\
 & (MJD) & (keV) & \multicolumn{2}{c}{(${10}^{-10} ~ \rm{ erg ~ {s}^{-1} ~ {cm}^{-2} }$)}
 & & \multicolumn{1}{c}{(MJD)} & \multicolumn{1}{c}{(s)} & (${10}^{-11} ~ \rm{ s ~ {s}^{-1} }$) & \\
\hline
1978 Mar & 43596        & 0.7--60 & $24 \pm 3$                      & $25 \pm 3$     & & 43596.7 & 7.679190(26)  & $-4.55$\footnotemark[$\S$]          & 1  \\
1979 Feb & 43928--43946 & 2--10   & $5.1 \pm 0.3$                   & $19 \pm 1$     & & 43946.0 & 7.677632(13)  & $-4.9 \pm 0.1$\footnotemark[$\S$]   & 2  \\
1983 May & 45457--45459 & 2--20   & $10.1 \pm 0.2$                  & $19.0 \pm 0.4$ & & 45458.0 & 7.671350(1)   & $-13 \pm 5$                         & 3  \\
1983 Aug & 45576        & 2--10   & $5.6 \pm 0.5$                   & $20.3 \pm 1.8$ & & 45576.9 & 7.67077(1)    & $-5.65 \pm 0.10$\footnotemark[$\l$] & 4  \\
1986 Mar & 46519--46520 & 1--20   & $14 \pm 0.4$\footnotemark[$\#$] & $22 \pm 0.7$   & & 46520.0 & 7.6664220(5)  & $-4.96 \pm 0.06$\footnotemark[$\S$] & 5  \\
1987 Mar & 46855--48012 & 1--20   & $8.89 \pm 0.56$                 & $14.0 \pm 0.9$ & &         &               &                                     & 6  \\
1988 Aug &              &         &                                 &                & & 47400.3 & 7.6625685(30) &                                     & 7  \\
1990 Apr & 47999--48002 & 2--60   & $\sim 20.6$                     & $\sim 24.7$    & & 48001.1 & 7.660069(2)   &                                     & 8  \\
1990 Jun & 48043--48530 & 1--20   & $6.67 \pm 0.89$                 & $10.9 \pm 1.5$ & &         &               &                                     & 6  \\
1990 Aug &              &         &                                 &                & & 48133.5 & 7.66001(4)    &                                     & 4  \\
1993 Aug & 49210--49211 & 0.5--10 & $\sim 2.8$                      & $\sim 9.9$     & &         &               &                                     & 9  \\
1996 Aug & 50301--50306 & 2--60   & $\sim 6.1$                      & $\sim 6.5$     & &         &               &                                     & 10 \\
2000 Sep & 51803        & 0.3--10 & 2.4\footnotemark[$**$]          & 8.4            & & 51803.6 & 7.6726(2)     &                                     & 11 \\
2001 Aug & 52145        & 0.3--10 & 1.8\footnotemark[$**$]          & 6.3            & & 52145.1 & 7.6736(2)     & $3.39 \pm 0.96$\footnotemark[$\l$]  & 11 \\
2003 Jun & 52793        & 0.3--10 & 1.8\footnotemark[$**$]          & 6.3            & & 52795.1 & 7.67514(5)    & $2.74 \pm 0.37$\footnotemark[$\l$]  & 11 \\
2003 Aug & 52871        & 0.3--10 & 1.6\footnotemark[$**$]          & 5.6            & & 52871.2 & 7.67544(6)    & $4.6 \pm 1.2$\footnotemark[$\l$]    & 11 \\
2007 Jun & 54280        & 15--50  & $2.8 \pm 0.1$                   & $5.6 \pm 0.2$  & & 54280   & $\sim 7.6793$ & 2.9                                 & 12 \\
2007 Sep & 54370        & 15--50  & $2.6 \pm 0.1$                   & $5.2 \pm 0.2$  & & 54370   & $\sim 7.6793$ & 2.7                                 & 12 \\
2007 Dec & 54450        & 15--50  & $4.0 \pm 0.1$                   & $8.0 \pm 0.3$  & & 54450   & $\sim 7.6793$ & 1.9                                 & 12 \\
2008 Jan & 54480        & 15--50  & $4.6 \pm 0.2$                   & $9.3 \pm 0.4$  & & 54480   & $\sim 7.6793$ & 0.18                                & 12 \\
2008 Feb & 54510        & 15--50  & $4.6 \pm 0.1$                   & $11.5 \pm 0.3$ & & 54510   & $\sim 7.6793$ & $-0.53$                             & 12 \\
2008 Mar & 54530        & 2--100  & $10.1 \pm 0.8$                  & $11.7 \pm 0.9$ & &         &               &                                     & 13 \\
2008 Mar & 54550        & 15--50  & $5.8 \pm 0.1$                   & $14.5 \pm 0.3$ & & 54550   & $\sim 7.6793$ & $-2.3$                              & 12 \\
2008 Jun & 54620        & 15--50  & $5.9 \pm 0.1$                   & $14.8 \pm 0.2$ & & 54620   & $\sim 7.6793$ & $-2.7$                              & 12 \\
\hline
\end{tabular}
\end{center}
\footnotesize
\par\noindent
All errors represent 1-$\sigma$ uncertainties.
\par\noindent
\footnotemark[$*$] Converted 0.5--100 keV flux, assuming the spectral models in \citet{2012A&A...546A..40C}.
\par\noindent
\footnotemark[$\dagger$] Values in parentheses are 1 $\sigma$ error in the last digit(s).
\par\noindent
\footnotemark[$\ddagger$] (1) \citet{1979ApJ...231..912P} (HEAO 1/A-2), (2) \citet{1983ApJ...266..769E} (Einstein/MPC),
 (3) \citet{1986PASJ...38..751K} (Tenma), (4) \citet{1994A&A...285..503M} (EXOSAT/GSPC, ROSAT), (5) \citet{1988ApJ...327..732L} (EXOSAT/ME),
 (6) \citet{1997astro.ph..7105V} (Ginga/ASM), (7) \citet{1990PASJ...42L..27S} (Ginga), (8) \citet{1995PhDT.......215M} (Ginga),
 (9) \citet{1995ApJ...449L..41A} (ASCA), (10) \citet{1998ApJ...500L.163O} (BeppoSAX), (11) \citet{2007ApJ...660..605K} (Chandra, XMM-Newton),
 (12) The data were read from figure 4 in \citet{2010ApJ...708.1500C} (Swift/BAT), (13) \citet{2010ApJ...708.1500C} (RXTE/PCA spectra).
\par\noindent
\footnotemark[$\S$] Averaged value.
\par\noindent
\footnotemark[$\l$] Averaged $\dot{P}$ calculated from $P$ of the observation and the previous one in this table.
\par\noindent
\footnotemark[$\#$] The error was estimated by the count rate with ME, using the values of the count rate and the error with GSPC.
\par\noindent
\footnotemark[$**$] Absorption-corrected flux.
\end{table*}

\footnotetext{http://gammaray.nsstc.nasa.gov/batse/pulsar/data/sources/4u1626.html}

\section{Data Analysis}

\begin{figure}
\begin{center}
\includegraphics[width=90mm,angle=-90]{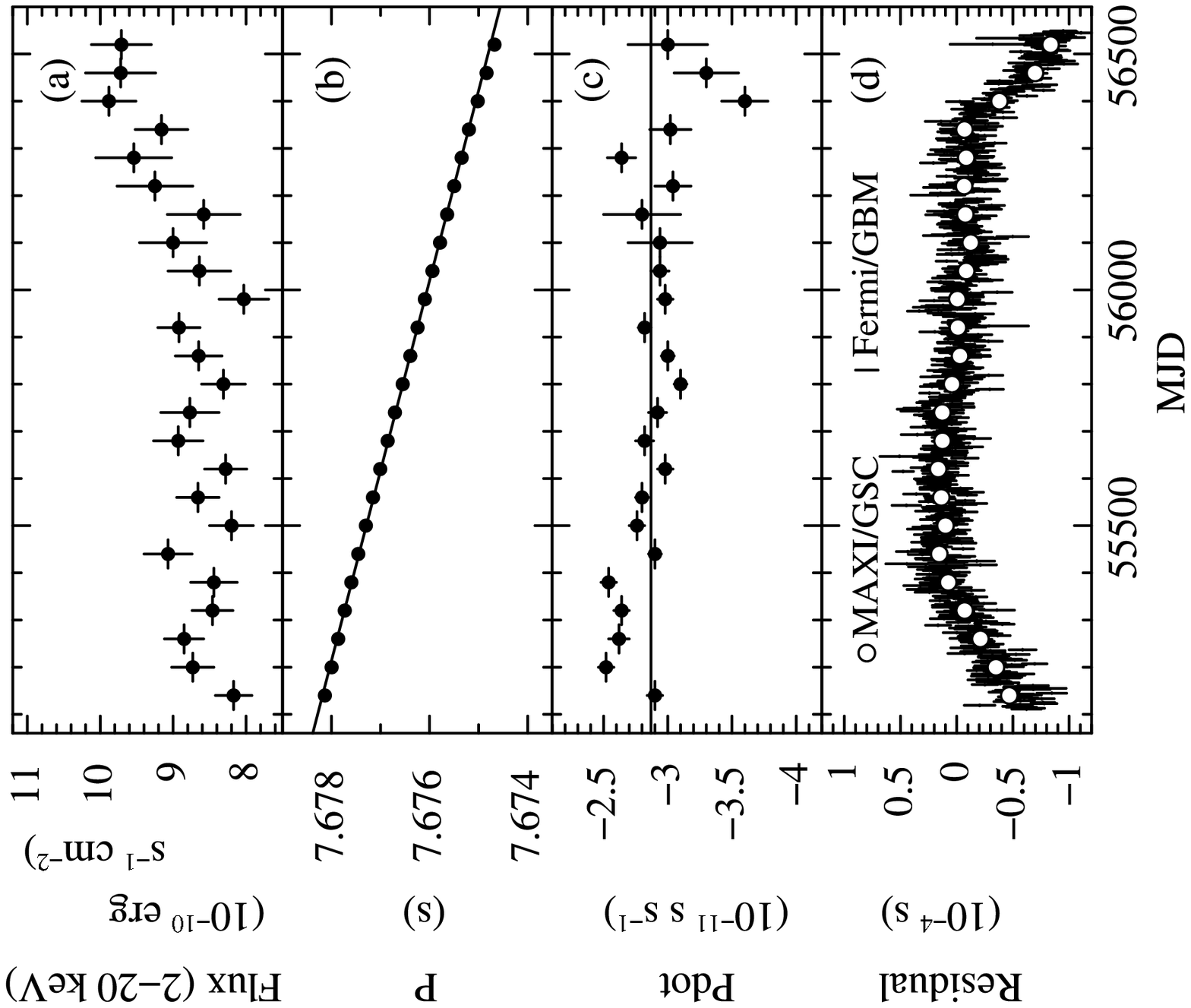}
\end{center}
\caption{The 2--20 keV flux (panel a), the period (panel b), and the period derivative (panel c) of 4U 1626--67,
 obtained every 60 d with the MAXI/GSC data from 2009 October to 2013 September.
Panel (d) shows residuals from the best-fit linear function to the period data in (b).
The horizontal line in (c) represents the slope of the line in (b) ($-2.87 \times {10}^{-11} ~ \rm{ s ~ {s}^{-1} }$).
The Fermi/GBM results are superposed by vertical line segments.}
\label{fig:maxi_fx_p_pdot}
\end{figure}

\subsection{X-ray light curve with MAXI}
X-ray events of 4U 1626--67 were extracted from all-sky GSC data, and accumulated over every 60-d interval from MJD 55110 (2009 October 6) to MJD 56550 (2013 September 15),
 using the on-demand analysis system provided by the MAXI team\footnote{http://maxi.riken.jp/mxondem}.
We employed the standard regions to extract the on-source and background events; ${2}^{\circ}$ radius circle for the source region
 and an annulus with inner and outer radii of ${2}^{\circ} .1$ and ${3}^{\circ}$, respectively, for the background.
We fitted all the obtained spectra with a power-law model without absorption.
The fits were acceptable for all the 60-d intervals, and gave photon indices of $1.0 \sim 1.3$.
We calculated the 2--20 keV model flux, and present in figure \ref{fig:maxi_fx_p_pdot}a its variation over the 4 years analyzed,
 where errors refer to 1$\sigma$ statistical uncertainties.
The flux was thus almost constant at $\sim 8.6 \times {10}^{-10} ~ \rm{ erg ~ {cm}^{-2} ~ {s}^{-1} }$ before MJD 56200,
 and slightly increased to $\sim 9.5 \times {10}^{-10} ~ \rm{ erg ~ {cm}^{-2} ~ {s}^{-1} }$ after that time.

\subsection{Pulse periods and pulse-period derivatives with MAXI}
\begin{figure}
\begin{center}
\includegraphics[width=65mm,angle=-90]{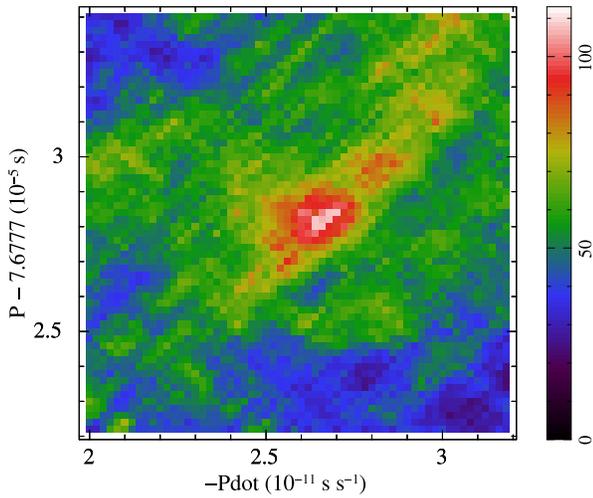}
\end{center}
\caption{Distribution of ${\chi}^{2}$ of the folded pulse profiles, shown as a function of the trial $P$ and $\dot{P}$,
 obtained from the 2--10 keV MAXI/GSC event data in MJD 55290--55350.
The right bar indicates the ${\chi}^{2}$ values.
The maximum ${\chi}^{2}$ is 108 for 31 degrees of freedom at $P = 7.6777282 ~ \rm{s}$ and $\dot{P} = -2.64 \times {10}^{-11} ~ \rm{ s ~ {s}^{-1} }$.}
\label{fig:cont_k}
\end{figure}

The pulsar timing analysis of 4U 1626--67 was carried out with the GSC event data of revision 1.5, which have a time resolution of 50 $\mu$s.
We extracted events within a ${1}^{\circ} .5$ radius from the source position, and then applied the barycentric correction to their arrival times.
Background was not subtracted in the timing analysis.
To determine $P$ and $\dot{P}$, we employed the epoch folding method.
There, ${\chi}^{2}$ of the folded pulse profile, defined as
\begin{equation}
{\chi}^{2} ~=~ \sum_{i = 1}^{n} {\left[ \frac{ ( {y}_{ \rm{i} } - \bar{y} ) }{ \sqrt{ {y}_{ \rm{i} } } } \right]}^{2},
 ~ \bar{y} ~=~ \frac{ \sum {( \frac{1}{ { \sqrt{ {y}_{ \rm{i} } } } } )}^{2} {y}_{ \rm{i} } }{ {\sum {( \frac{1}{ { \sqrt{ { {y} }_{ \rm{i} } } } } }) }^{2} },
\label{equ:chi_s}
\end{equation}
 is calculated for each trial value of $P$ and $\dot{P}$, where $n$ is the number of bins of the folded profile and ${y}_{ \rm{i} }$ is the number of events in the $i$-th bin.
We used $n = 32$, and confirmed that the exposure is uniform over the 32 bins within 0.5\%.
The epoch-folding analysis was performed for every 60-d interval, to be coincident with the light-curve time bins employed in section 2.1.
In each interval, we searched for the values of $P$ and $\dot{P}$ that maximize ${\chi}^{2}$.
Here, the $P$ and $\dot{P}$ values measured with the Fermi/GBM pulsar project\footnote{http://gammaray.nsstc.nasa.gov/gbm/science/\\pulsars/lightcurves/4u1626.html}
 were used to select the search ranges, and $\dot{P}$ was assumed constant in each interval.
Figure \ref{fig:cont_k} shows the obtained ${\chi}^{2}$ values on the $P$--$\dot{P}$ plane, employing MJD 55290--55350 as a typical example.
The 1-$\sigma$ errors of $P$ and $\dot{P}$ were estimated by Monte-Carlo simulations (see Appendix 3).
We repeated the analysis in the energy bands of 2--20 keV, 2--10 keV, 2--4 keV, 4--10 keV and 10--20 keV,
 and then selected the results of the 2--10 keV band because the maximum ${\chi}^{2}$ was the highest among them.
Results from the other energy bands were consistent with these.

Figure \ref{fig:maxi_fx_p_pdot}b and \ref{fig:maxi_fx_p_pdot}c show time variations of the obtained $P$ and $\dot{P}$, respectively.
The absolute value of $\dot{P}$ increased with the flux increase around MJD 56400.
We fitted the data in figure \ref{fig:maxi_fx_p_pdot}b with a liner function, because their distribution appears quite linear.
The best-fit slope was then obtained as $\langle \dot{P} \rangle = -2.87 \times {10}^{-11} ~ { \rm{ s~{s}^{-1} } }$.
Figure \ref{fig:maxi_fx_p_pdot}d shows the residuals from the best-fit line,
 where the results of the Fermi/GBM pulsar data are overlaid.
The results of the MAXI/GSC and the Fermi/GBM are found to agree with each other within the errors.

\subsection{Estimation of bolometric flux}
In the following sections, we apply the theoretical models of pulsar spin-up/down,
 proposed by GL79 and LRB95, to the observed data including those from the previous measurements and from the MAXI/GSC.
For this, we have to estimate the bolometric flux ${F}_{ \rm{bol} }$ of the individual observations,
 considering different energy bands used in different observations, and employing appropriate spectral models.

The energy spectrum of 4U 1626-67 were studied in both the spin-up and spin-down phases (table \ref{tab:past_fx_p_pdot}),
 and the changes in the spectral shape between these two phases were reported (e.g. \cite{2010MNRAS.403..920J}; \cite{2012A&A...546A..40C}).
According to \citet{2012A&A...546A..40C}, the spectra in both phases can be fitted with a model composed of a blackbody and a cutoff power law,
 and their difference is in the blackbody component, whose temperature is $\sim$ 0.5 keV in the spin-up phase and $\sim$ 0.2 keV in the spin-down phase.
We hence employed these respective spectral models for the spin-up and spin-down phases,
 and converted the 2--20 keV MAXI/GSC flux to those in the 0.5--100 keV band, which we identify with ${F}_{ \rm{bol} }$.
Since the power-law continuum is flat (photon index $\sim$ 1) below $\sim$ 20 keV and exponentially cuts off above $\sim$ 20 keV,
 the fluxes in the energy band below 0.5 keV and above 100 keV are negligible.
For example, the conversion factor from the 2--20 keV flux observed by the MAXI/GSC (in the spin-up phase) to the 0.5--100 keV flux is 1.88.
In a similar way, we calculated ${F}_{ \rm{bol} }$ in the past observations, and present the results in table \ref{tab:past_fx_p_pdot}.

\begin{figure*}
\begin{center}
\includegraphics[width=130mm,angle=-90]{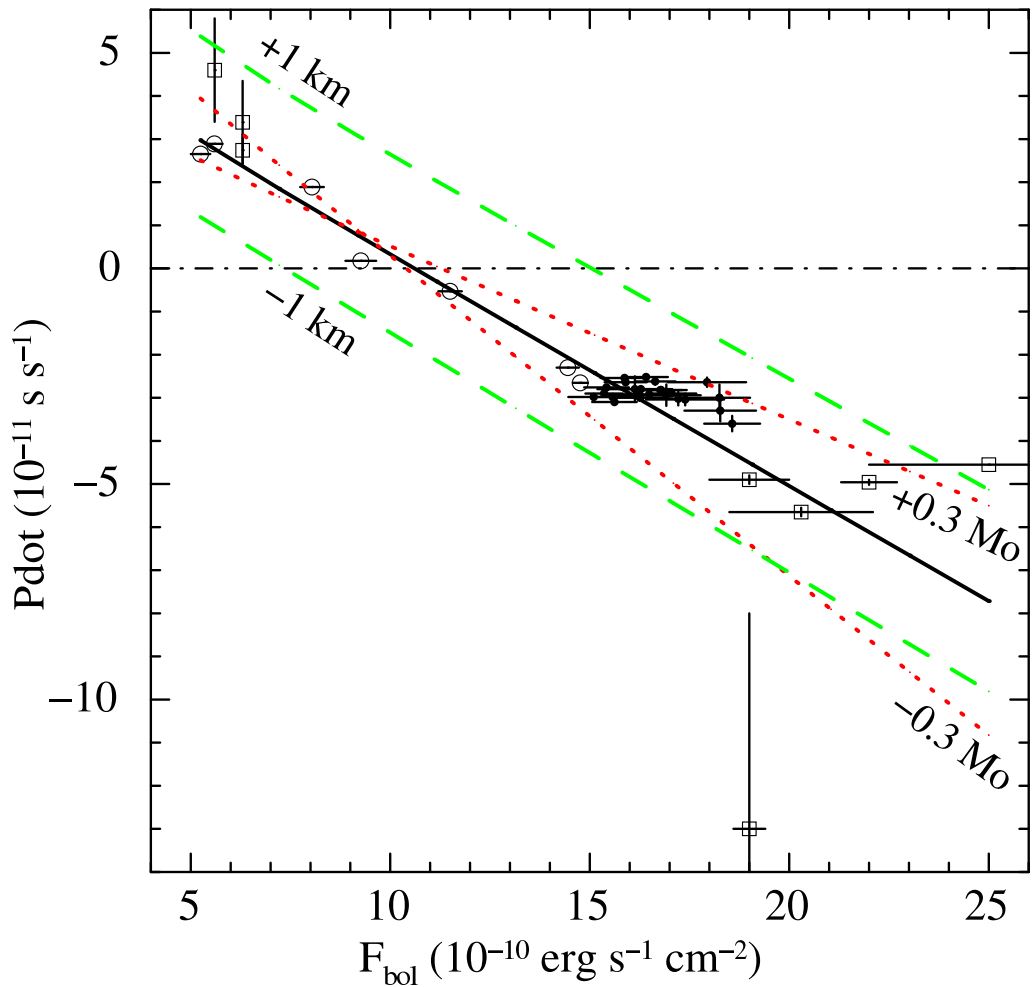}
\end{center}
\caption{A relation between $\dot{P}$ and ${F}_{ \rm{bol} }$ by the MAXI/GSC and the past observations.
Filled circles, open circles and squares represent the MAXI/GSC data, the Swift/BAT data and the others in table \ref{tab:past_fx_p_pdot}, respectively.
The dashed-dotted horizontal line indicates $\dot{P} = 0$.
The solid line is the best fit model calculated by equation (\ref{equ:GandL}), assuming a distance of 10 kpc.
The parameters are $M = 1.83 {M}_{\odot}$ and $R = 11.4$ km, with ${ {\chi}_{\nu} }^{2}$ of 2.9 for 37 degrees of freedom.
Dashed two lines show the case when $R$ is changed by $\pm 1$ km with $M$ kept unchanged,
 while dotted two lines those when $M$ is varied by $\pm 0.3 ~ {M}_{\odot}$ with $R$ fixed at 11.4 km.}
\label{fig:lx_flux_gl}
\end{figure*}

In figure \ref{fig:lx_flux_gl}, we plot the relation between the observed $\dot{P}$ and the calculated ${F}_{ \rm{bol} }$, including the past data.
It clearly shows their negative correlation, expected from the pulsar spin up due to the accretion torque.
Furthermore, the data points in the spin-up and spin-down phases apparently defines a well-defined single dependence on ${F}_{ \rm{bol} }$.

\section{Application of the Ghosh \& Lamb relation}

As reviewed in section 1, GL79 derived a relation between $\dot{P}$ and $L$ in accreting X-ray pulsars,
 assuming that the accreting matter transfers its angular momentum to the pulsar at the ``outer transition zone'', ${r}_{ \rm{0} }$ [equation (\ref{equ:rA})].
The equations we used are summarized in Appendix 1.
\citet{1977Natur.266..683R} also proposed an almost equivalent equation.
Since their model equation includes an unknown parameter ($\xi {v}_{ \rm{r} } / {v}_{ \rm{ff} }$),
 of which the relation to $n({\omega}_{ \rm{s} })$ in the GL79 relation is calculable.
Therefore, we employ the GL79 relation.

\subsection{Model equations relating $\dot{P}$ to the flux}
According to GL79, $\dot{P}$ is expressed by
\begin{equation}
\dot{P} ~=~ - 5.0 \times {10}^{-5} { {\mu}_{30} }^{ \frac{2}{7} } n({\omega}_{ \rm{s} })
 {S}_{1}(M) {P}^{2} { {L}_{37} }^{ \frac{6}{7} } ~ { \rm{ s ~ {yr}^{-1} } },
\label{equ:GandL}
\end{equation}
where ${\mu}_{30}$ is the magnetic dipole moment $\mu$ in units of ${10}^{30} ~ { \rm{ G~{cm}^{3} } }$,
 and ${L}_{37}$ is $L$ in units of ${10}^{37} ~ { \rm{ erg~{s}^{-1} } }$.
The functions $n({\omega}_{ \rm{s} })$ and ${S}_{1}(M)$ are given in Appendix 1,
 where ${\omega}_{\rm{s}}$ is the fastness parameter defined by equation (\ref{equ:omega_s}).
If ${\omega}_{\rm{s}}$ is in the range of $0 - 0.9$, $n({\omega}_{ \rm{s} })$ is approximated by equation (\ref{equ:n_omega_s}) within 5\%.
Since $n({\omega}_{ \rm{s} })$ changes from positive to negative depending on $\omega_{ \rm{s} }$ (figure \ref{fig:n_omega_s}),
 $\dot{P}$ can become both positive ($\omega_{ \rm{s} } > 0.349$; spin down) and negative ($\omega_{ \rm{s} } < 0.349$; spin up).

As shown in equation (\ref{equ:s1}), ${S}_{1}(M)$ contains the effective moment of inertia $I$.
It is expressed as a function of $M$ and $R$, dependent on the EOS.
We utilize its approximation given by \citet{2005ApJ...629..979L},
\begin{eqnarray}
I ~{\simeq}~ && (0.237 \pm 0.008) M {R}^{2} \nonumber \\
 \times && \left[1 + 0.42 \left( \frac{M}{ {M}_{\odot} } \right) \left( \frac{R}{ 10 ~ {\rm{km}} } \right)^{-1}
 + 0.009 \left( \frac{M}{ {M}_{\odot} } \right)^{4} \left( \frac{R}{ 10 ~ {\rm{km}} } \right)^{-4} \right], \nonumber \\
\label{equ:moment}
\end{eqnarray}
which is applicable in most of the major EOS models if $M / R \gtrsim 0.07 {M}_{\odot} ~ { \rm{ {km}^{-1} } }$.

In 4U 1626--67, the surface magnetic field is known from the cyclotron resonance scattering feature as
  ${B}_{ \rm{c} } = 3.2 \times {10}^{12} ~ (1 + {z}_{ \rm{g} }) ~ { \rm{G} }$ (section 1).
It is considered to represent the field strength near the magnetic poles.
Assuming that the magnetic axis is aligned to the pulsar rotation axis, $\mu$ at the equator in the GL79 model is represented by
\begin{equation}
\mu ~=~ \frac{1}{2} {B}_{ \rm{c} } {R}^{3}.
\label{equ:mu}
\end{equation}

If the source emission is isotropic, $L$ is calculated from ${F}_{ \rm{bol} }$ and the distance $D$ as
\begin{equation}
L ~=~ 4 \pi {D}^{2} {F}_{ \rm{bol} }.
\label{equ:luminosity}
\end{equation}
Since the pulsar emission is anisotropic, it is not exactly correct.
We employ this approximation and then discuss the effect later.

Combining equations (\ref{equ:GandL}), (\ref{equ:moment}), (\ref{equ:mu}), and (\ref{equ:luminosity}),
 as well as the expression for ${z}_{ \rm{g} }$ [equation (\ref{equ:z_g})], we obtain a theoretical model equation to describe the observed $\dot{P}$--$L$ relation,
 including three unknown parameters, $D$, $M$ and $R$.

\subsection{Comparison between the data and theory}
In order to determine the possible parameter ranges of $D$, $M$ and $R$,
 we fitted the observed $\dot{P}$--${F}_{ \rm{bol} }$ relation in figure \ref{fig:lx_flux_gl} with the model prepared as above.
When errors associated with some past measurements of $\dot{P}$ are unavailable,
 we assumed an arbitrarily small value ($\Delta \dot{P} = 6 \times {10}^{-16} ~ { \rm{s ~ {s}^{-1} } }$),
 because the overall errors are dominated by those in ${F}_{ \rm{bol} }$.
This treatment was confirmed to little affect the fitting results.
Since it is difficult to constrain all the three parameters simultaneously from the $\dot{P}$--${F}_{ \rm{bol} }$ relation alone,
 we first assumed the source distance $D$ to be some values from 3 to 20 kpc, and then determined the allowed $M$--$R$ regions as a function of the assumed distance.
As an example, the fitting result assuming $D = 10$ kpc is shown in figure \ref{fig:lx_flux_gl},
 where the best fit values were obtained as $M = 1.83 ~ {M}_{\odot}$ and $R = 11.4$ km (errors are considered later).

To understand how the model curve depends on $M$ and $R$,
 we show in figure \ref{fig:lx_flux_gl} some predictions by equation (\ref{equ:GandL}) when either $M$ or $R$ is slightly changed.
Thus, changes in $R$ (with $D$ and $M$ fixed) causes parallel displacements of the model with little changes in its slope,
 while those in $M$ (with $D$ and $R$ fixed) appears mainly in slope changes with the ``zero-cross'' point not much affected.
In other words, the observed $\dot{P}$--$F_{\rm bol }$ relation has essentially two degrees of freedom, namely, the zero-cross point and the slope,
 and their joint use allows us to simultaneously constrain two (in the present case $M$ and $R$) out of the three model parameters:
 the other one ($D$ in the present case) remains unconstrained.
Below, let us consider physical meanings behind this model behavior.

In figure \ref{fig:lx_flux_gl}, the zero-cross point at the spin-up/down threshold represents a torque-equilibrium condition,
 wherein so-called co-rotation radius, which is almost uniquely determined by the observed $P$ (with some dependence on $M$),
 can be equated with magnetospheric radius, or the outer-transition radius ${r}_{0}$ in the GL79 model (Appendix 1).
At this radius, the gravitational pressure calculated from $L \propto {D}^{2} {F}_{ \rm{bol} }$ should balance the magnetic pressure,
 and this condition specifies the value of $\mu$.
By further comparing this $\mu$ with the accurately measured ${B}_{ \rm{c} }$, we can constrain $R$ via equation (\ref{equ:mu}).
As a result, the zero-cross point becomes more sensitively to $R$ rather than to $M$.
More quantitatively, the torque-equilibrium condition in equation (\ref{equ:omega_s}), ${\omega}_{ \rm{s} } = 0.349$, can be combined with equation (\ref{equ:mu}),
 to yield a scaling for the flux at the torque equilibrium as
\begin{equation}
{F}_{ \rm{bol} } \propto {M}^{ - \frac{2}{3} } {R}^{5} {D}^{-2} ,
\label{equ:phy_f_bol}
\end{equation}
where dependences on ${B}_{ \rm{c} }$ and $P$ were omitted.

The slope of the $\dot{P}$--${F}_{ \rm{bol} }$ relation in figure \ref{fig:lx_flux_gl} means
 the conversion factor from an increment of the luminosity (and hence of the accretion torque) to that in the neuron-star rotation.
Thus, it is inversely proportional to $I$, so that an increase in $M$ will make the slope smaller (in the absolute value).
A larger $R$ will act in the same sense through $I$, but this effect is partially canceled by an induced increase in $\mu$,
 through equation (\ref{equ:GandL}), which would make period changes easier.
As a result, the slope becomes mainly determined by $M$.
Quantitatively, at the highest spin-up regime which most accurately specifies the slope, we can approximate $\omega \rightarrow 0$,
 and hence $n({\omega}_{ \rm{s} }) \sim$ constant from figure \ref{fig:n_omega_s}, to rewrite equation (\ref{equ:GandL}) as
\begin{equation}
- \dot{P} \propto {M}^{ - \frac{10}{7} } {R}^{ - \frac{2}{7} } {L}^{ \frac{6}{7} }
 \propto {M}^{ - \frac{10}{7} } {R}^{ - \frac{2}{7} } {D}^{ \frac{12}{7} } { {F}_{ \rm{bol} } }^{ \frac{6}{7} }
\label{equ:phy_pdot}
\end{equation}
when ignoring the higher-order terms in equation (\ref{equ:moment}).
This yields the slope as
\begin{equation}
- \frac{d \dot{P} }{d {F}_{ \rm{bol} } } \propto {M}^{ - \frac{10}{7} } {R}^{ - \frac{2}{7} } {D}^{ \frac{12}{7} } { {F}_{ \rm{bol} } }^{ - \frac{1}{7} } .
\label{equ:phy_dpdot_dfbol}
\end{equation}

By changing the assumed $D$, we calculated the best-fit $M$ and $R$, and show their locus as a solid line in figure \ref{fig:mass_radi_dis},
 where the mass-radius relations from representative EOSs
 [SLy \citep{2001A&A...380..151D}, APR \citep{1998PhRvC..58.1804A} and Shen (\cite{1998NuPhA.637..435S}; \cite{1998PThPh.100.1013S}) presented in \citet{2013PhRvD..88b3009Y}]
 are also shown.
In order for the derived $M$ and $R$ fall in the nominal neutron-star parameters range,
 $M = (1.0 - 2.4) ~ {M}_{\odot}$ and $R = 8.5 - 15$ km (e.g. \cite{2010AdSpR..45..949B, 2013RPPh...76a6901O}), the distance must be $D = 5 - 13$ kpc.
This is in a good agreement with \citet{1998ApJ...492..342C}.
For reference, the locus in figure \ref{fig:mass_radi_dis} can be analytically calculated in the following way.
When a value of $D$ is given, the measured zero-point flux specifies ${M}^{-2/3} {R}^{5} {D}^{-2}$ via equation (\ref{equ:phy_f_bol}),
 while the slope in figure \ref{fig:lx_flux_gl} specifies ${M}^{-10/7} {R}^{-2/7} {D}^{12/7} {F}_{ \rm{bol} }^{-1/7}$ via equation (\ref{equ:phy_dpdot_dfbol}).
By eliminating $D$ from these two scalings, and ignoring the weakly varying factor ${F}_{ \rm{bol} }^{-1/7}$, we obtain
\begin{equation}
M \propto {R}^{2}
\label{equ:phy_mass}
\end{equation}
which approximately agrees with the locus in figure \ref{fig:mass_radi_dis}.

In figure \ref{fig:lx_flux_gl}, the best-fit reduced chi-squared,
 ${ {\chi}_{\nu} }^{2} = 2.9$ for $\nu = 37$ degrees of freedom (DOF), is not within the acceptable range.
One possible cause for this large ${\chi}^{2}$ may be systematic errors on the observed fluxes,
 because the flux data taken from various past results must be subject to cross-calibration uncertainties among the different instruments employed.
We thus repeated the model fitting by gradually increasing the systematic errors in the flux from 1\%,
 to find that ${ {\chi}_{\nu} }^{2} \sim 1$ is attained if the systematic errors are set at 5.5\%.
This number is quite reasonable, because the fluxes of the Crab nebula measured by these instruments scatter by $\sim 10\%$ \citep{2005SPIE.5898...22K}
 most likely due to uncertainties in the absolute photometric sensitivities of these instruments.

Since the fit ${ {\chi}_{\nu} }^{2}$ was found to depend little on $D$, we have decided to employ the systematic error of 5.5\% throughout,
 and calculated the statistically allowed $M$--$R$ region at the 68\% confidence limits (${\chi}^{2}$ increment $\Delta {\chi}^{2} < 2.3$ for 2 DOF).
In figure \ref{fig:mass_radi_dis}, the obtained allowed region is indicated by a pair of dashed lines,
 and the ranges of $M$ and $R$ for representative distances of 6, 7, 8,..., 13 kpc are listed in table \ref{tab:mass_radi_ns_fitting}.

\subsection{Systematic uncertainties}
\label{sec:GL79uncertainty}
\begin{figure}
\begin{center}
\includegraphics[width=70mm,angle=-90]{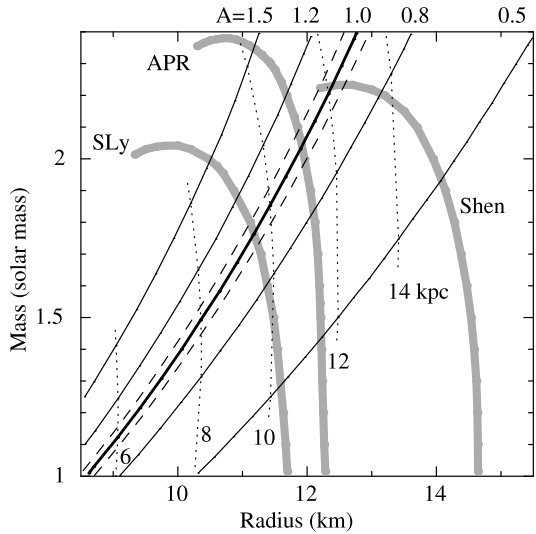}
\end{center}
\caption{Various constraints on the neutron star parameters, shown on the mass-radius plane.
The solid lines indicate how the best-fit values of $M$ and $R$, allowed by the data in figure \ref{fig:lx_flux_gl}, vary as $D$ is changed.
The cases of four different values of the normalization factor $A$ are shown.
A pair of dashed lines represent 68\% errors for the $A = 1.0$ curve, while dotted lines give contours of the source distance to 4U 1626--67.
The gray solid lines are mass-radius relations predicted by three EOSs; SLy, APR and Shen.}
\label{fig:mass_radi_dis}
\end{figure}

Although the present model fitting has been found to have a capability of rather accurately constraining $M$ and $R$ when $D$ is given,
 the uncertainty range in figure \ref{fig:mass_radi_dis} (dashed lines) considers only statistical errors.
We therefore need to evaluate systematic errors associated with several assumptions and approximations which we have employed.
Among them, the approximations involved in equation (\ref{equ:n_omega_s}) for $n({\omega}_{ \rm{s} })$ of the GL79 model,
 and equation (\ref{equ:moment}) for $I$, are considered to be accurate to within $< 5 \%$ (GL79) and $< 10 \%$ (\cite{2005ApJ...629..979L}), respectively.
We hence neglecting these effects, and consider below more dominant sources of systematic errors.

One obvious uncertainty is in equation (\ref{equ:luminosity}),
 which assumes that the time-averaged flux of a pulsar is identical to the average over the whole direction.
Although the difference between these two averages has not been estimated in 4U 1626--67,
 \citet{1975A&A....42..311B} examined this issue in Her X-1, a similar low-mass X-ray binary pulsar, and concluded that the difference is at most 50\%.
Assuming that the condition is similar in 4U 1626--67, we assign a systematic uncertainty to the flux up to $\sim 50\%$,
 which is transferred almost directly to that in the normalization factor of equation (\ref{equ:GandL}).
Another uncertainty inherent to the GL79 model is those in the accretion geometry,
 including the exact location of the ``outer transition zone'' radius ${r}_{0}$ of equation (\ref{equ:rA}),
 and the angles among the pulsar's rotation axis, its magnetic axis, and the accretion plane;
 we assumed that the rotation and magnetic axes are parallel, and are perpendicular to the accretion plane.
All these effects may be represented effectively by uncertainty in ${\mu}_{30}$.
Because ${r}_{0}$ is proportional to ${ {\mu}_{30} }^{4/7}$ and the right hand side of equation (\ref{equ:GandL}) to ${ {\mu}_{30} }^{2/7}$,
 an uncertainty in ${\mu}_{30}$ by, e.g., 50\%, would induce a 25\% change in the coefficient of equation (\ref{equ:GandL}).

To jointly take into account all these uncertainties, we have decided to introduce an artificial normalization factor $A$,
 and multiplied it to the right hand side of equation (\ref{equ:GandL}).
Then, the model fitting was repeated by changing $A$ from 0.5 to 1.5.
In figure \ref{fig:mass_radi_dis}, the allowed $M$--$R$ regions for $A =$ 0.5, 0.8, 1.2, and 1.5 are also drawn.
Thus, the uncertainty indeed affects the mass determination, but $R$ remains very well constrained as long as $D$ is somehow determined.

\begin{table}
\caption{Allowed $M$--$R$ regions of the neutron star of 4U 1626--67 for an assumed distance corresponding to the dashed lines in figure \ref{fig:mass_radi_dis}}
\label{tab:mass_radi_ns_fitting}
\begin{center}
\begin{tabular}{cccc}
\hline
Assumed distance & Mass\footnotemark[$*$] & & Radius\footnotemark[$*$] \\
(kpc) & (${M}_{\odot}$) & & (km) \\
\hline
6.0  & 1.09--1.15 & & 9.03--9.11 \\
7.0  & 1.28--1.34 & & 9.70--9.79 \\
8.0  & 1.45--1.53 & & 10.3--10.4 \\
9.0  & 1.63--1.71 & & 10.9--11.0 \\
10.0 & 1.81--1.90 & & 11.4--11.5 \\
11.0 & 1.98--2.08 & & 11.8--12.0 \\
12.0 & 2.15--2.26 & & 12.2--12.4 \\
13.0 & 2.32--2.43 & & 12.6--12.8 \\
\hline
\end{tabular}
\end{center}
\footnotesize
\par\noindent
\footnotemark[$*$] 68\% confidence (2-parameters errors) limits.
\end{table}

\section{Application of the Lovelace model}
As described in section 1 and detailed in Appendix 2, LRB95 developed another (in a sense more sophisticated) model,
 to be called ``Lovelace model'' here, to explain the relation between $\dot{P}$ and $L$, assuming magnetic outflows and/or magnetic breaking.
Using the ``turnover radius'' ${r}_{ \rm{to} }$ [equation (\ref{equ:r_to})] and the co-rotation radius ${r}_{ \rm{cr} }$ [equation (\ref{equ:r_cr})],
 the model predicts spin-up with outflows when ${r}_{ \rm{to} } < {r}_{ \rm{cr} }$,
 and provides both spin-up and spin-down solutions with magnetic braking by the disk when ${r}_{ \rm{to} } > {r}_{ \rm{cr} }$.
To explain both the spin-up and spin-down behavior of 4U 1626--67 with the LRB95 model, ${r}_{ \rm{to} } > {r}_{ \rm{cr} }$ must therefore be satisfied in the spin-down phase.
However, in the spin-down phase of 4U 1626--67, we found that ${r}_{ \rm{to} } > {r}_{ \rm{cr} }$ is not satisfied under the parameters which we assumed.
For example, the values are estimated to be ${r}_{ \rm{to} } = 1.3 \times {10}^{8}$ cm and ${r}_{ \rm{cr} } = 7.0 \times {10}^{8}$ cm
 with ${F}_{ \rm{bol} } = 9.3 \times {10}^{-10} ~ \rm{ erg ~ {cm}^{-2} ~ {s}^{-1} }$, $M = 1.73 {M}_{\odot}$, $R = 11.1$ km, $D = 6$ kpc, and $\alpha {D}_{ \rm{m} } = 0.01$.
Thus, the Lovelace model cannot explain the spin-down phase of 4U 1626--67.

Even though the Lovelace model has the above problem, it might provide a reasonable explanation to the spin-up behavior of 4U 1626--67.
Because ${r}_{ \rm{to} } < {r}_{ \rm{cr} }$ is satisfied in the spin-up phase of this object, we should employ the ``spin-up with outflows'' solution by LRB95,
which describes the $\dot{P}$--$L$ relation as
\begin{eqnarray}
\dot{P} ~&{\approx}&~ -4.3 \times {10}^{-5} { {\mu}_{30} }^{0.285}{ \left( \frac{ \alpha {D}_{ \rm{m} } }{0.1} \right) }^{0.15} { {R}_{6} }^{0.85} \nonumber \\
&& \times { \left( \frac{M}{ {M}_{\odot} } \right) }^{-0.425} { {I}_{45} }^{-1} {P}^{2} { {L}_{37} }^{0.85} ~ { \rm{ s ~ {yr}^{-1} } },
\label{equ:Lovelace_Lx}
\end{eqnarray}
where $\alpha$ is the viscous parameter in \citet{1973A&A....24..337S}, ${D}_{ \rm{m} }$ is the magnetic diffusivity parameter,
 ${R}_{6}$ is $R$ in units of ${10}^{6} ~ \rm{cm}$, and ${I}_{45}$ is $I$ in units of ${10}^{45} ~ \rm{ g ~ {cm}^{2} }$.
In LRB95, $\alpha$ is assumed to be 0.01 to 0.1 and ${D}_{ \rm{m} }$ to be of order of unity.
Equation (\ref{equ:Lovelace_Lx}) is equivalent to equation (\ref{equ:GandL}),
 where the the major difference is that the factor $n({\omega}_{ \rm{s} })$ in the latter is replaced by $\alpha {D}_{ \rm{m} }$ in the former.

\begin{figure}
\begin{center}
\includegraphics[width=70mm,angle=-90]{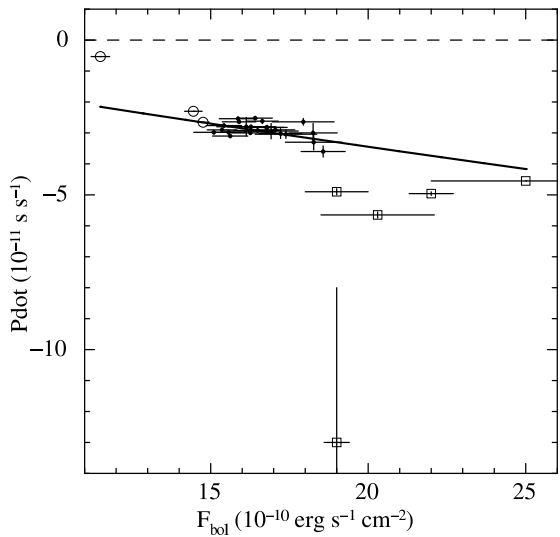}
\end{center}
\caption{A scatter plot between $\dot{P}$ and ${F}_{ \rm{bol} }$ in the spin-up phases,
 obtained by the MAXI/GSC and other satellites, presented in the same manner as figure \ref{fig:lx_flux_gl}.
The dashed horizontal line indicates $\dot{P} = 0$.
The solid line is the best fit model by equation (\ref{equ:Lovelace_Lx}), with $M = 1.73 {M}_{\odot}$, $R = 11.1$ km, $D = 6$ kpc, and $\alpha {D}_{ \rm{m} } = 0.01$.
The fit goodness is ${ {\chi}_{\nu} }^{2} = 46$ (1400/30) for 30 DOF.}
\label{fig:lx_flux_ll}
\end{figure}

We thus selected the spin-up-phase data from table \ref{tab:past_fx_p_pdot}, and fitted them with equation (\ref{equ:Lovelace_Lx}),
 over the parameter ranges of $M = 1.0 - 2.4 {M}_{\odot}$ and $R = 8.5 -15$ km.
A result for $D = 6$ kpc and $\alpha {D}_{ \rm{m} } = 0.01$ is shown in figure \ref{fig:lx_flux_ll},
 in the same format as figure \ref{fig:lx_flux_gl} (but limited to $\dot{P} < 0$).
The fit is far from being acceptable, with ${ {\chi}_{\nu} }^{2} = 46$ for 30 DOF.
Changing $D$ or $\alpha {D}_{ \rm{m} }$ did not solve the problem.
This is not surprising, since equation (\ref{equ:Lovelace_Lx}) can explain a torque-equilibrium condition ($\dot{P} = 0$) only when the flux tends to zero.
This make a contrast to the GL79 model, and disagrees with the measurements.
In conclusion, the LRB95 model cannot explain the observed behavior of 4U 1626--67.

\section{Discussion}
Applying the epoch folding analysis to the MAXI/GSC data,
 we determined $P$, $\dot{P}$, and the X-ray flux of 4U 1626--67 for every 60-d interval from 2009 October to 2013 September.
The pulsar has been spinning up throughout this period, and the spin-up rate was positively correlated with the flux.

On the $\dot{P}$--${F}_{ \rm{bol} }$ plane (figure \ref{fig:lx_flux_gl}),
 these MAXI/GSC results were confirmed to be fully consistent with those from the past observations (table \ref{tab:past_fx_p_pdot}).
In fact, the overall data jointly define a well defined $\dot{P}$ vs. ${F}_{ \rm{bol} }$ correlation,
 covering rather evenly the spin-up and spin-down phases from $\dot{P} = - 6 \times {10}^{-11} ~ \rm{ s ~ {s}^{-1} }$ to $+ 5 \times {10}^{-11} ~ \rm{ s ~ {s}^{-1} }$.
Utilizing these favorable conditions,
 we have successfully shown that the accretion-torque theory by GL79 can adequately explain the overall $\dot{P}$--${F}_{ \rm{bol} }$ behavior of 4U 1626--67,
 while that of LRB95 model cannot explain the data even limiting the comparison to the spin-up phase.

Another favorable condition of this object is the accurate knowledge of its surface magnetic field.
As a result, the observed $\dot{P}$--${F}_{ \rm{bol} }$ relation were found to constrain, via the GL79 model, two of the three unknown parameters;
 the distance $D$, the mass $M$ of the pulsar, and its radius $R$.
When $M$ and $R$ are allowed to take any value in the nominal mass and radius ranges of neutron stars,
 namely $M = (1.0 - 2.4) ~ {M}_{\odot}$ and $R = 8.5 - 15$ km respectively, the distance can be constrained to $D = 5 - 13$ kpc.
This is consistent with the limit $D \gtrsim$ 3 kpc derived by \citet{1997ApJ...474..414C};
 these authors analyzed the $\dot{P}$ behavior of 4U 1626--67 under the assumption of $I = 1 \times {10}^{45} ~ \rm{ g ~ {cm}^{2} }$ and $M = 1.4 ~ {M}_{\odot}$,
 then estimated the mass accretion rate to be $\dot{M} \gtrsim 1 \times {10}^{16} ~ \rm{ g ~ {s}^{-1} }$ using a similar consideration to the GL79 framework,
 and compared the expected luminosity to the observed flux.
Our distance estimate is also consistent with that of \citet{1998ApJ...492..342C}, 5--13 kpc,
 which was derived from the optical flux assuming that the effective X-ray albedo of the accretion disk is $\gtrsim 0.9$.

Conversely, if the distance $D$ is assumed, the $\dot{P}$--${F}_{ \rm{bol} }$ relation can fix $M$ and $R$ with relatively small statistical errors.
Table \ref{tab:mass_radi_ns_fitting} lists the allowed $M$ and $R$ ranges for typical source distances assumed.
Thus, once $D$ can be determined by some other means, $M$ and $R$ of 4U 1626--67 can be constrained to a rather narrow range,
 which would be useful to pin-down the nuclear EOS.
A point of particular importance is that the present method can provide the information on $R$,
 which is more vitally needed than that on $M$, without not much affected by various systematic errors (figure \ref{fig:mass_radi_dis}).

In order to further increase the reliability of the GL79 method, it is important to understand the systematic errors (section \ref{sec:GL79uncertainty}).
In this respect, an application of the GL79 model to Be X-ray binaries by \citet{2014MNRAS.437.3863K} is worth noting.
They compared the surface magnetic field of these pulsars calculated using the GL79 model ($ = {B}_{ \rm{GL79} }$),
 with that measured using cyclotron resonance scattering features ($ = {B}_{ \rm{C} }$).
The ratio was found as ${B}_{ \rm{GL79} } / {B}_{ \rm{C} } = 3 - 4$ in two examples, GRO J1008--57 and A0535+26.
This corresponds to a value of $A \simeq 1.5$, which is consistent with the $50\%$ uncertainty in $A$ assumed in section \ref{sec:GL79uncertainty}.
Thus, studying a larger number of sources would be important.

\begin{ack}
The authors are grateful to all members of the MAXI team, BATSE pulsar team and Fermi/GBM pulsar project.
This work was supported by RIKEN Junior Research Associate Program,
 and the Ministry of Education, Culture, Sports, Science and Technology (MEXT), Grant-in-Aid No. 24340041.
\end{ack}

\appendix

\section{Ghosh \& Lamb model}
GL79 derived an equation between $\dot{P}$ and $L$ in an X-ray binary pulsar.
The accreting matter transfers the angular momentum to the pulsar at the ``outer transition zone'', ${r}_{ \rm{0} }$.
Equation (11) in GL79 denotes ${r}_{ \rm{0} } = 0.52\ {r}_{ \rm{A} }^{ \rm{(0)} }$,
 where ${r}_{ \rm{A} }^{ \rm{(0)} }$ is the characteristic Alfven radius.
Substituting the numbers
\begin{equation}
{r}_{ \rm{A} }^{ \rm{(0)} } ~=~ 3.2 \times {10}^{8} ~ { { \dot{M} }_{17} }^{ - \frac{2}{7} } ~ { {\mu}_{30} }^{ \frac{4}{7} }
 ~ { \left( \frac{M}{ { {M}_{\odot} } } \right) }^{ - \frac{1}{7} } ~ \rm{cm}, 
\label{equ:rA0}
\end{equation}
\begin{equation}
{r}_{ \rm{0} } ~=~ 1.7 \times {10}^{8} ~ { { \dot{M} }_{17} }^{ - \frac{2}{7} } ~ { {\mu}_{30} }^{ \frac{4}{7} }
 ~ { \left( \frac{M}{ { {M}_{\odot} } } \right) }^{ - \frac{1}{7} } ~ \rm{cm}, 
\label{equ:rA}
\end{equation}
where ${ \dot{M} }_{17}$ is the accretion rate $\dot{M}$ in units of ${10}^{17} ~ { \rm{ g ~ {s}^{-1} } }$.

\begin{figure}
\begin{center}
\includegraphics[width=60mm,angle=-90]{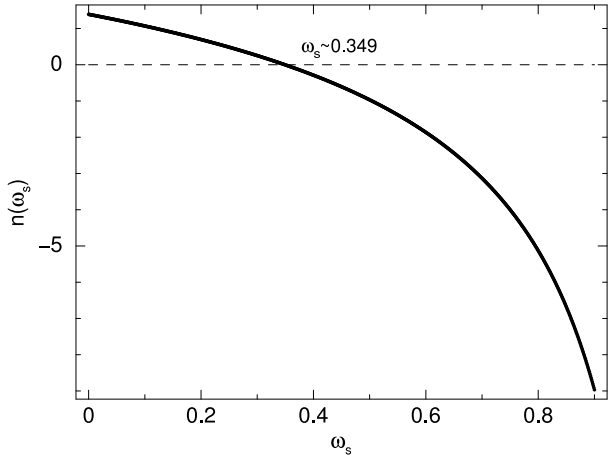}
\end{center}
\caption{The function $n({\omega}_{ \rm{s} })$ with ${\omega}_{ \rm{s} }$.
The approximation by equation (\ref{equ:n_omega_s}) is effective in ${\omega}_{ \rm{s} } = 0 - 0.9$.
Thus, $n({\omega}_{ \rm{s} })$ becomes 0 at ${\omega}_{ \rm{s} } \sim 0.349$, is positive in ${\omega}_{ \rm{s} } \lesssim 0.349$,
 and negative in ${\omega}_{ \rm{s} } \gtrsim 0.349$.}
\label{fig:n_omega_s}
\end{figure}

GL79 gave their theoretical $\dot{P}$--$L$ relation [equation (15) in GL79] as
\begin{equation}
\dot{P} ~=~ - 5.0 \times {10}^{-5} { {\mu}_{30} }^{ \frac{2}{7} } n({\omega}_{ \rm{s} })
 {S}_{1}(M) {P}^{2} { {L}_{37} }^{ \frac{6}{7} } ~ \rm{ s ~ {yr}^{-1} },
\label{equ:GandLApp}
\end{equation}
where $L$ is defined by $L ~ = ~ \dot{M} (GM / R)$.
The functions $n({\omega}_{ \rm{s} })$ and ${S}_{1}(M)$ are described respectively by equations (10) and (17) in GL79 as
\begin{eqnarray}
n({\omega}_{ \rm{s} }) ~&{\approx}&~ 1.39 [1 - {\omega}_{ \rm{s} } \{4.03 {(1 - {\omega}_{ \rm{s} } )}^{0.173} - 0.878\}] \nonumber \\
&& \times { (1 - {\omega}_{ \rm{s} }) }^{-1},
\label{equ:n_omega_s}
\end{eqnarray}
\begin{equation}
{S}_{1}(M) ~=~ { {R}_{6} }^{ \frac{6}{7} } { \left( \frac{M}{ {M}_{\odot} } \right) }^{ - \frac{3}{7} } { {I}_{45} }^{-1}.
\label{equ:s1}
\end{equation}
Here, ${\omega}_{ \rm{s} }$ is the so-called fastness parameter, which is a dimensionless parameter,
 defined as the ratio of the pulsar's angular frequency to the Keplerian angular frequency of the accreting matter.
This ${\omega}_{ \rm{s} }$ is expressed approximately by equation (16) in GL79 as
\begin{equation}
{\omega}_{ \rm{s} } ~{\approx}~ 1.35 ~ { {\mu}_{30} }^{ \frac{6}{7} } ~ {S}_{2}(M) ~ {P}^{-1} { {L}_{37} }^{ - \frac{3}{7} }.
\label{equ:omega_s}
\end{equation}
Here, ${S}_{2}(M)$ is given by equation (18) in GL79 as
\begin{equation}
{S}_{2}(M) ~=~ { {R}_{6} }^{ - \frac{3}{7} } { \left( \frac{M}{ {M}_{\odot} } \right) }^{ - \frac{2}{7} }.
\label{equ:s2}
\end{equation}
Equation (\ref{equ:n_omega_s}) is accurate to within 5\% for $0 \leq {\omega}_{ \rm{s} } \leq 0.9$.
The behavior of $n({\omega}_{ \rm{s} })$ is plotted in figure \ref{fig:n_omega_s}, where the zero crossover point is seen to take place at ${\omega}_{ \rm{s} } \sim 0.349$.

\section{Lovelace model}
LRB95 introduced magnetic outflows and magnetic breaking to explain the relation between $\dot{P}$ and $\dot{M}$.
After many numerical integrations they introduced ${r}_{ \rm{to} }$,
 where the angular velocity ${\omega}_{ \rm{a} }$ of accreting matter reaches the maximum (${ \rm{d} } {\omega}_{ \rm{a} } / { \rm{d} } r = 0$).
The matter transfers the angular momentum to the pulsar at $r = {r}_{ \rm{to} }$.
${r}_{ \rm{to} }$ is given by equation (16) in LRB95 as
\begin{equation}
{r}_{ \rm{to} } ~{\approx}~ 0.91 \times {10}^{8} { \left( \frac{ \alpha {D}_{ \rm{m} } }{0.1} \right) }^{0.3} {\mu}_{30}^{0.57}
 { \dot{M} }_{17}^{-0.3} { \left( \frac{M}{ { {M}_{\odot} } } \right) }^{-0.15} ~ { \rm{cm} }.
\label{equ:r_to}
\end{equation}
In LRB95, $\alpha$ is assumed as 0.01 to 0.1 and ${D}_{ \rm{m} }$ is order unity.
${r}_{ \rm{to} }$ has basically the same nature as ${r}_{ \rm{0} }$ in the GL79 model [equation (\ref{equ:rA})].
According to ${r}_{ \rm{to} }$ and ${r}_{ \rm{cr} }$,
 LRB95 demonstrates a magnetic outflow case (${r}_{ \rm{to} } < {r}_{ \rm{cr} }$)
 and magnetic braking of the disk case (${r}_{ \rm{to} } > {r}_{ \rm{cr} }$), where ${r}_{ \rm{cr} }$ is
\begin{equation}
{r}_{ \rm{cr} } ~{\equiv}~ { \left( \frac{GM}{ { {\omega}_{*} }^{2} } \right) }^{ \frac{1}{3} }
 {\approx}~ 1.5 \times {10}^{8} { \left( \frac{M}{ {M}_{\odot} } \right) }^{ \frac{1}{3} } {P}^{ \frac{2}{3} } ~ { \rm{cm} },
\label{equ:r_cr}
\end{equation}
and ${\omega}_{*} = 2 \pi / P$.

When ${r}_{ \rm{to} }$ is smaller than ${r}_{ \rm{cr} }$, the pulsar shows spin-up with magnetic outflow.
$\dot{P}$ equation [equation (18b) in LRB95] consists $\dot{M}$ and ${r}_{ \rm{to} }$.
By moving $P$ to the right side,
\begin{equation}
\dot{P} ~{\approx}~ -5.8 \times {10}^{-5} {P}^{2} { \dot{M} }_{17} {I}_{45}^{-1}
 { \left( \frac{M}{ {M}_{\odot} } \right) }^{ \frac{1}{2} } { \left( \frac{ {r}_{ \rm{to} } }{ {10}^{8} ~ { \rm{cm} } } \right) }^{ \frac{1}{2} }
 ~ { \rm{ s ~ {yr}^{-1} } }.
\label{equ:Lovelace}
\end{equation}
By deducing $\dot{M} = L R / G M$,
\begin{eqnarray}
\dot{P} ~&{\approx}&~ -4.3 \times {10}^{-5} { {\mu}_{30} }^{0.285} { \left( \frac{ \alpha {D}_{ \rm{m} } }{0.1} \right) }^{0.15} { {R}_{6} }^{0.85} \nonumber \\
&& \times { \left( \frac{M}{ { {M}_{\odot} } } \right) }^{-0.425} { {I}_{45} }^{-1} {P}^{2} { {L}_{37} }^{0.85} ~ { \rm{ s ~ {yr}^{-1} } }.
\label{equ:Lovelace_Lx_app}
\end{eqnarray}
It is equivalent to equation (\ref{equ:GandLApp}).
The indices are the same, and the factor is almost the same.
The difference is the $\alpha {D}_{ \rm{m} }$ term instead of the $n({\omega}_{ \rm{s} })$ term.

When ${r}_{ \rm{to} }$ is larger than ${r}_{ \rm{cr} }$, magnetic braking takes place.
It can work for both spin-up and spin-down, although it mostly works as spin-down.

\section{Error estimation of period and period derivative}
We usually use the folding method to obtain $P$ and $\dot{P}$ of the pulse, orbit etc.
First we assume a set of $P$ and $\dot{P}$, then fold the light curve with them.
If there is no periodicity the resultant folded light curve is flat.
If there is a pulsation with $P$ and $\dot{P}$ it shows a pulse shape.
We calculate ${\chi}^{2}$ which is a sum of squared deviation from the mean, to evaluate the existence of a pulse.
We change $P$ and $\dot{P}$ to find the most-likely $P$ and $\dot{P}$ which give the maximum ${\chi}^{2}$.
We can draw distribution of ${\chi}^{2}$ in the $P$ and $\dot{P}$ plane.
Figure \ref{fig:cont_k} is an example which we used in this paper to obtain $P$ and $\dot{P}$ of 4U 1626--67.
Thus we can derive $P$ and $\dot{P}$, however, their errors are not obvious.

\subsection{Method 1 - parameter $a$ and the standard estimate -}
The most primitive and straight-forward method for error-estimation is to assume that
 the errors $\Delta P, ~ \Delta \dot{P}$ corresponds to a difference of pulse number $a$
 in the whole time span ${T}_{ \rm{s} }$ of the observation.
\begin{equation}
\Delta P ~=~ \frac{ a {P}^{2} }{ {T}_{ \rm{s} } }
\hspace{15mm}
\Delta \dot{P} ~=~ \frac{ 2a {P}^{2} }{ { {T}_{ \rm{s} } }^{2} }
\label{equ:p_pdot_equ_1}
\end{equation}
$a$ is usually less than 1 pulse.
Equation (\ref{equ:p_pdot_equ_1}) are led by as follows.
The number of pulses $n(t)$ since ${t}_{0}$ is an integral of the pulse frequency $\nu(t)$ from ${t}_{0}$ to $t$.
Let us assume that the frequency $\nu$ depends on time with a constant rate $\dot{\nu}$,
 or we take only the first order of derivatives with time in Taylor expansion series.
Using $\nu(t') ~=~ {\nu}_{0} + { \dot{\nu} }_{0} (t' - {t}_{0})$,
\begin{equation}
n(t) ~=~ \int_{ {t}_{0} }^{t} \nu (t') ~ dt' ~=~ {\nu}_{0} (t - {t}_{0}) + \frac{ \dot{ {\nu}_{0} } }{2} {( t - {t}_{0})}^{2} .
\label{equ:pulse_number_1}
\end{equation}
Using the whole time span ${T}_{ \rm{s} } = (t - {t}_{0})$,
\begin{equation}
n(t) ~=~ {\nu}_{0} {T}_{ \rm{s} } + \frac{ \dot{ {\nu}_{0} } }{2} ~ {T}_{ \rm{s} }^{2}.
\label{equ:pulse_number_2}
\end{equation}
Equation (\ref{equ:pulse_number_2}) is also considered as a function of ${\nu}_{0}$ and $\dot{ {\nu}_{0} }$.
We can calculate the ``variation of $n(t)$'', $\Delta n$, as a function of variations of ${\Delta \nu}_{0}$ and $\Delta \dot{ {\nu}_{0} }$.
\begin{eqnarray}
\Delta n(\Delta {\nu}_{0}, \Delta \dot{ {\nu}_{0} })
&~=~& \frac{ \partial n(t) }{ \partial {\nu}_{0} } \Delta {\nu}_{0} + \frac{ \partial n(t) }{ \partial { \dot{\nu} }_{0} }{ \Delta { \dot{\nu} }_{0} } \nonumber \\
&~=~& {T}_{ \rm{s} } \Delta {\nu}_{0} + \frac{ { {T}_{ \rm{s} } }^{2} }{2} \Delta { \dot{\nu} }_{0}
\label{equ:delta_n}
\end{eqnarray}
In this section we defined $\Delta n ~ = ~ a$, and we obtain $\Delta {\nu}_{0}$ and $\Delta \dot{ {\nu}_{0} }$ from equation (\ref{equ:delta_n}).
\begin{equation}
\Delta {\nu}_{0} ~=~ \frac{a}{ {T}_{ \rm{s} } }
\hspace{12mm}
\Delta { \dot{\nu} }_{0} ~=~ \frac{2 a}{ { {T}_{ \rm{s} } }^{2} }
\label{equ:delta_nu_nudot_h}
\end{equation}
Converting $\Delta {\nu}_{0}$ and $\Delta \dot{ {\nu}_{0} }$ to $\Delta {P}_{0}$ and $\Delta \dot{ {P}_{0} }$ in equation (\ref{equ:delta_nu_nudot_h}),
\begin{equation}
\Delta {P}_{0} ~=~ \frac{ a { {P}_{0} }^{2} }{ T_{ \rm{s} } },
\hspace{7mm}
\Delta { \dot{P} }_{0} ~=~ \frac{ 2a { {P}_{0} }^{2} }{ { {T}_{ \rm{s} } }^{2} } + \frac{ 2a {P}_{0} { \dot{P} }_{0} }{ {T}_{ \rm{s} } }.
\label{equ:delta_p_pdot_h}
\end{equation}
The second term of $\Delta { \dot{P} }_{0}$ can be ignored for ${P}_{0} / {T}_{ \rm{s} } \gg { \dot{P} }_{0}$.
In our case, we can ignore the term.

In this appendix, we describe various methods with the $a$ value.
\citet{1987A&A...180..275L} called $a = 1/2$ as the ``standard estimate'' and gave
\begin{equation}
\frac{ \Delta P }{P} ~=~ \frac{P}{ 2 {T}_{ \rm{s} } } .
\label{equ:standard_estimate}
\end{equation}
When we use $a = 1/2$ in our case (${T}_{ \rm{s} } = 60 ~ { \rm{days} } ~ = 5184000 ~ { \rm{s} }$, and $P = 7.67 ~ { \rm{s} }$), the errors of $P$ and $\dot{P}$ are
\begin{equation}
\Delta P ~=~ 5.7 \times {10}^{-6} ~ { \rm{s} },
\hspace{5mm}
\Delta \dot{P} ~=~ 2.2 \times {10}^{-12} ~ { \rm{ s ~ {s}^{-1} } }.
\label{equ:err_p_pdot_1}
\end{equation}
However, there is no reason to choose $a = 1/2$.

\subsection{Method 2 - sinusoidal pulse -}
It seems natural to consider that $a$ should be related to the reduced chi-square (${ {\chi}_{\nu} }^{2}$) value of the folded light curve.
Through the Monte-Carlo simulation, \citet{1987A&A...180..275L} obtained an empirical relation between $a$ and ${ {\chi}_{\nu} }^{2}$ for a sinusoidal pulse shape as
\begin{equation}
\frac{ \Delta P }{ \Delta {P}_{ \rm{L} } } ~=~ 0.71 ~ {({ {\chi}_{ \rm{\nu} } }^{2} - 1)}^{-0.63},
\label{equ:leahy_sigma_p}
\end{equation}
where $\Delta {P}_{ \rm{L} } = {P}^{2} / 2 {T}_{ \rm{s} }$.
Or, in our notation,
\begin{equation}
a ~=~ \frac{1}{2} \times 0.71 ~ {({ {\chi}_{ \rm{\nu} } }^{2} - 1)}^{-0.63}.
\label{equ:leahy_a}
\end{equation}
The equation is valid within ${ {\chi}_{\nu} }^{2} = 3 - 110$ which they investigated.
In 4U 1626--67, ${ {\chi}_{\nu} }^{2} = 3.48$ and we can use equation (\ref{equ:leahy_a}).
Then equation (\ref{equ:leahy_a}) gives $a = 0.20$.
By using equation (\ref{equ:delta_p_pdot_h}), the error of $P$ is 
\begin{equation}
\Delta P ~=~ 2.3 \times {10}^{-6} ~ \rm{s}.
\label{equ:err_p_2a}
\end{equation}
The error of $\dot{P}$ was not given in \citet{1987A&A...180..275L}.
However, if we assume that $a$ value by equation (\ref{equ:leahy_a}) might also be effective to calculate the error of $\dot{P}$,
\begin{equation}
\Delta \dot{P} ~=~ 0.88 \times {10}^{-12} ~ \rm{ s ~ {s}^{-1} }.
\label{equ:err_pdot_2a}
\end{equation}

\subsection{Method 2 modified - considering the pulse shape -}

\begin{figure}
\begin{center}
\includegraphics[width=50mm,angle=-90]{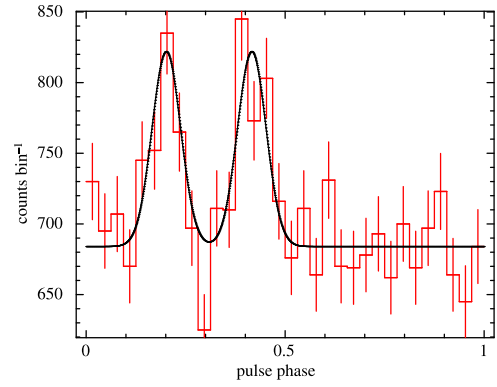}
\end{center}
\caption{Observed pulse shape in MJD 55290--55350 and a pulse model in 2--10 keV band.
The model consists two gaussian peaks and a constant.
Two gaussians have an equal height and an equal width ($\sigma = 0.036$, or FWHM = 0.085).
${ {\chi}_{\nu} }^{2} = 1.30$.
The background rate is 440 counts/bin.}
\label{fig:pulse_shape_model}
\end{figure}

In method 2, the pulse shape is expressed by a sine function.
This can be considered as the worst case among various pulse shapes, since it has the broadest pulse width as wide as 0.5 (FWHM) in phase.
In 4U 1626--67, however, the pulse width is as sharp as 0.085 in phase (figure \ref{fig:pulse_shape_model}).
Sharper the pulse shape is, better the period would be determined.
Therefore, the error of 4U 1626--67 would also become 0.085/0.5 of that in method 2 ($a = 0.20 \times 0.085 / 0.5 = 0.034$).
Thus 
\begin{equation}
\Delta P ~=~ 0.39 \times {10}^{-6} ~ { \rm{s} },
\hspace{5mm}
\Delta \dot{P} ~=~ 0.15 \times {10}^{-12} ~ { \rm{ s ~ {s}^{-1} } }.
\label{equ:err_p_pdot_2b}
\end{equation}
Here we ignore the effect that the pulse shape has two peaks.

\subsection{Method 3 - deviation from the best pulse profile -}
\begin{figure}
\begin{center}
\includegraphics[width=65mm,angle=-90]{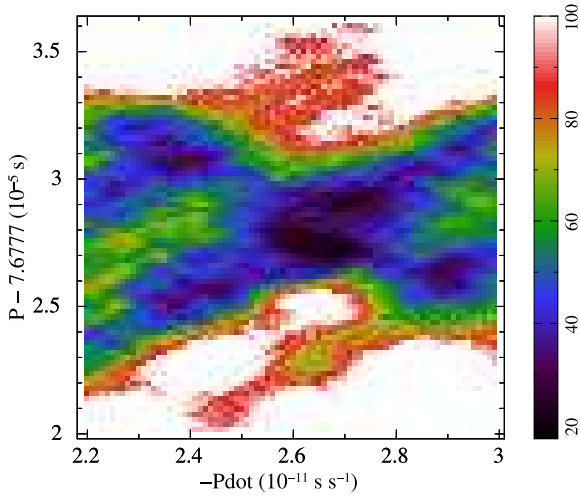}
\end{center}
\caption{Distribution of ${\chi}^{2}$ which represents deviation for the best pulse model (Method 3).
The right bar indicates the ${\chi}^{2}$ values.
The minimum of ${\chi}^{2}$ is 17.48.}
\label{fig:cont_k_chi_minimum}
\end{figure}

If we establish a good fit-model to the pulse profile and ${ {\chi}_{\nu} }^{2}$ is about 1.0,
 we could apply chi-square method to obtain errors of $P$ and $\dot{P}$.
We should note that ${\chi}^{2}$ of the fitting to the folded light curve is different from
 ${\chi}^{2}$ of the fitting of the observed light curve by repeating pulse shape model.
First, we determine the best-fit model for the folded pulse shape as figure \ref{fig:pulse_shape_model}.
Let us make the model to fit the data acceptably well, and fix it.
Then, we vary $P$ to the point where the fit of the model to the data is no longer acceptable.
We take the difference from the best-fit as an error, $\Delta P$.
Likewise for $\dot{P}$.
Figure \ref{fig:cont_k_chi_minimum} shows a distribution of ${\chi}^{2}$.
Here ${\chi}^{2}$ has the minimum around the center and it becomes larger as it goes apart.
By using the area where ${\chi}^{2}$ is the minimum plus 1.0, we obtained the one-parameter error of each $P$ and $\dot{P}$.
\begin{equation}
\Delta P ~=~ 0.2 \times {10}^{-6} ~ { \rm{s} }
\hspace{5mm}
\Delta \dot{P} ~=~ 0.4 \times {10}^{-12} ~ { \rm{ s ~ {s}^{-1} } }
\label{equ:err_p_pdot_3}
\end{equation}
We calculate back that $a$ is 0.02 and 0.09 for $P$ and $\dot{P}$, respectively.

\subsection{Method 4 - Monte-Carlo simulation -}
\begin{figure}
\begin{center}
\includegraphics[width=65mm,angle=-90]{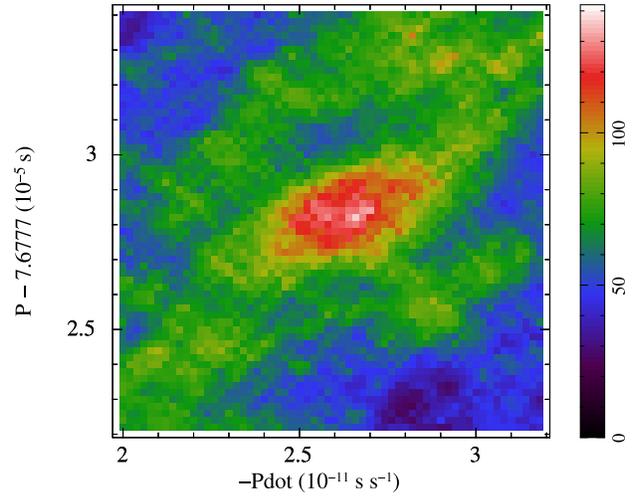}
\end{center}
\caption{Distribution of ${\chi}^{2}$ of folded pulse profiles on trial $P$ and $\dot{P}$,
 calculated for simulated event data in the same observation condition as that in MJD 55290--55350.
The right bar indicates the ${\chi}^{2}$ values.
The maximum ${\chi}^{2}$ is 137 for 31 degrees of freedom at $P = 7.6777282$ s and $\dot{P} = -2.66 \times {10}^{-11} ~ \rm{ s ~ {s}^{-1} }$.}
\label{fig:cont_k_sim}
\end{figure}

\begin{figure}
\includegraphics[width=30mm,angle=-90]{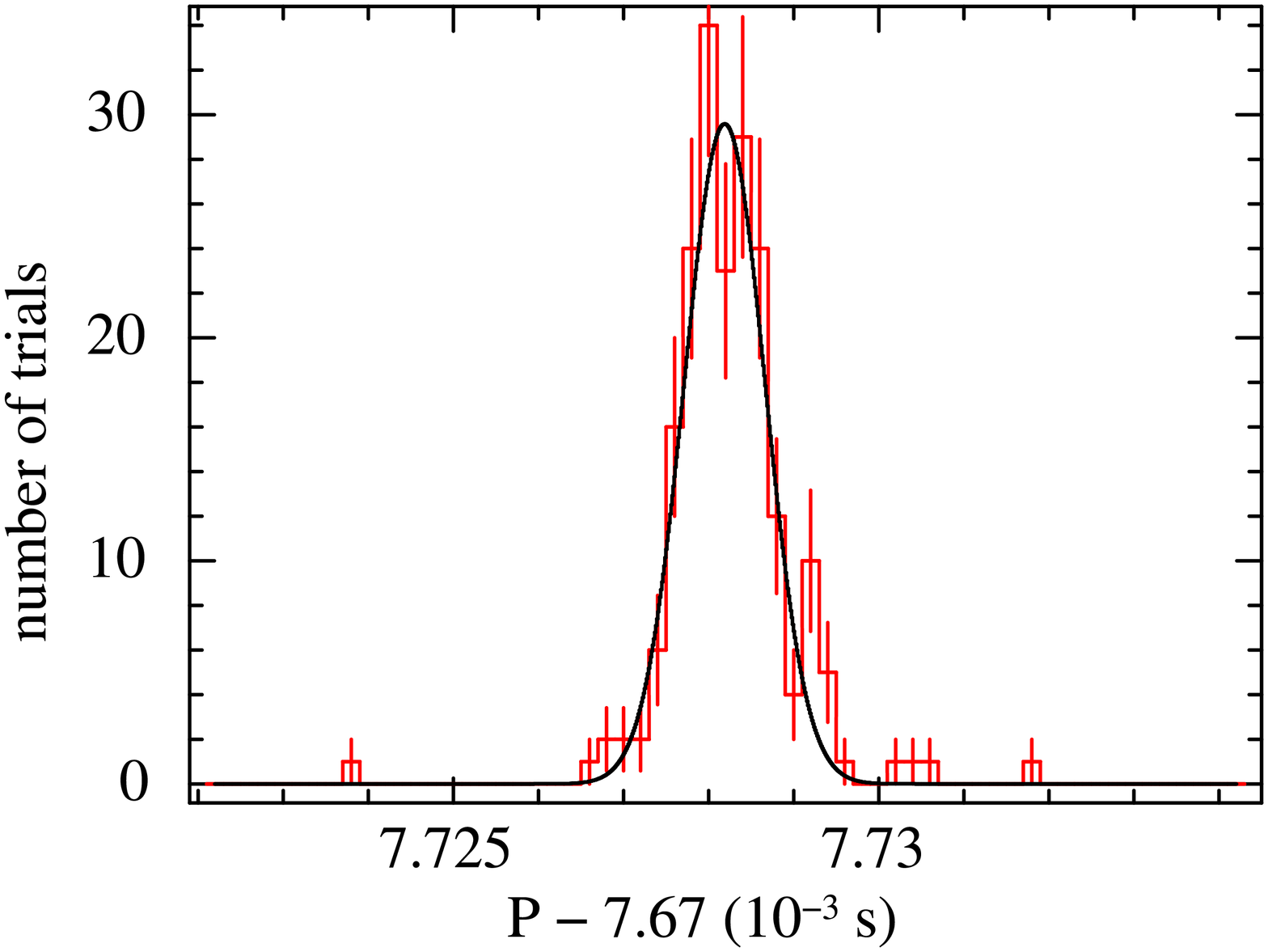}
\includegraphics[width=30mm,angle=-90]{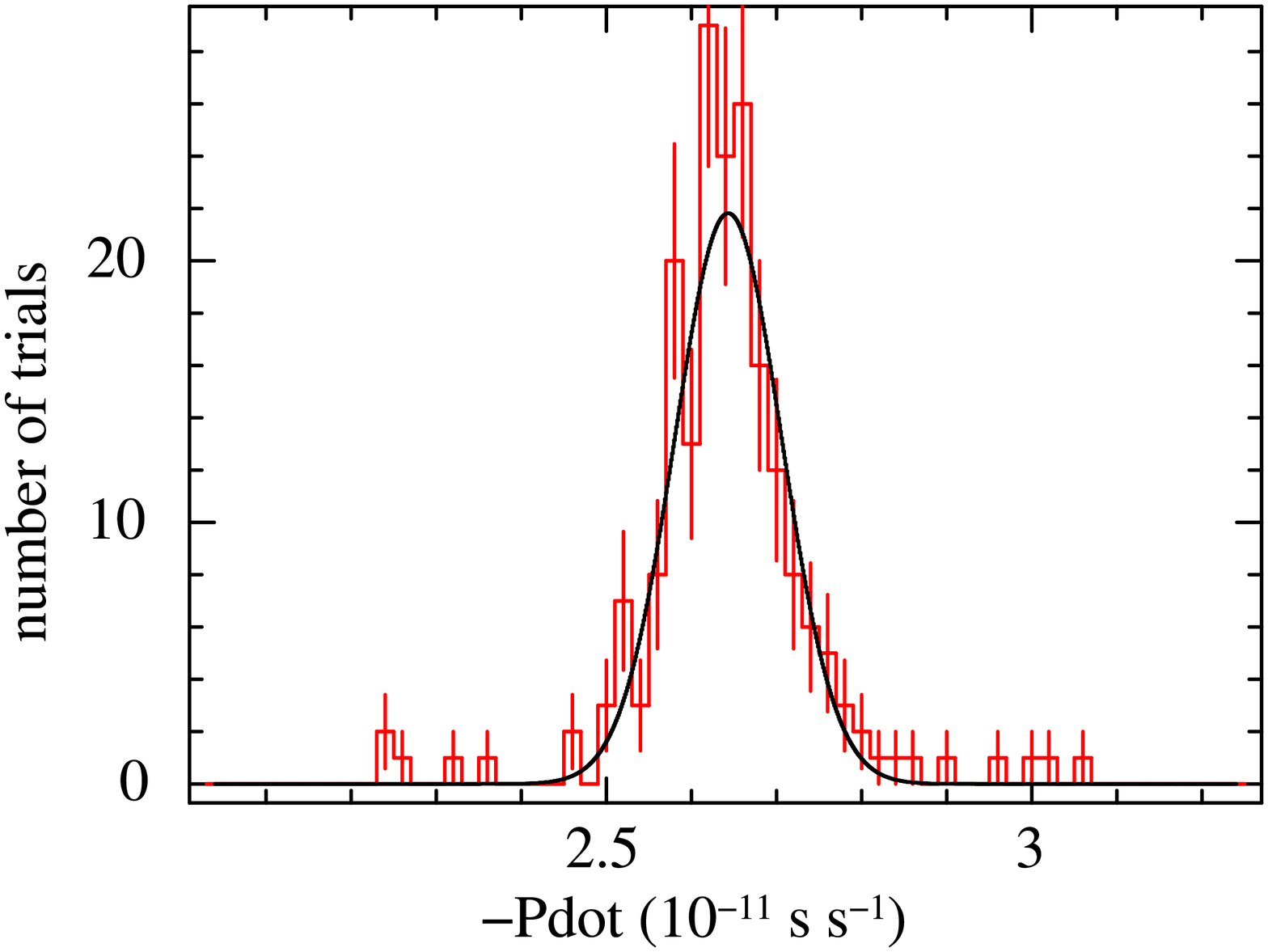}
\caption{Histogram of $P$ (left) and $\dot{P}$ (right).
The number of trials is 200.}
\label{fig:peri_pdot_hist}
\end{figure}

We carry out a Monte-Carlo simulation for the X-ray photons and the background taking into account all the observational conditions,
 such as source intensity, background intensity, accumulated area, exposure, and times of the scans.
Using simulated events, we obtain the most-likely $P$ and $\dot{P}$ by the folding method just as we did for the real observation.
Figure \ref{fig:cont_k_sim} shows an example of distribution of ${\chi}^{2}$ obtained by simulated data.
The ${\chi}^{2}$ values and the extension are similar to the real case (figure \ref{fig:cont_k}).
Then we repeat it many ($\sim 100$) times and make a histogram of each resultant $P$ and $\dot{P}$ in figure \ref{fig:peri_pdot_hist}.
The errors of $P$ and $\dot{P}$ are given by the gaussian widths (1 $\sigma$) of the histograms.
\begin{equation}
\Delta P ~=~ 0.48 \times {10}^{-6} ~ { \rm{s} }
\hspace{5mm}
\Delta \dot{P} ~=~ 0.63 \times {10}^{-12} ~ { \rm{ s ~ {s}^{-1} } }
\label{equ:err_p_pdot_4}
\end{equation}

\noindent
We calculate back that $a$ is 0.042 and 0.14 for $P$ and $\dot{P}$, respectively.

\subsection{Discussion}
\begin{figure}
\begin{center}
\includegraphics[width=50mm,angle=-90]{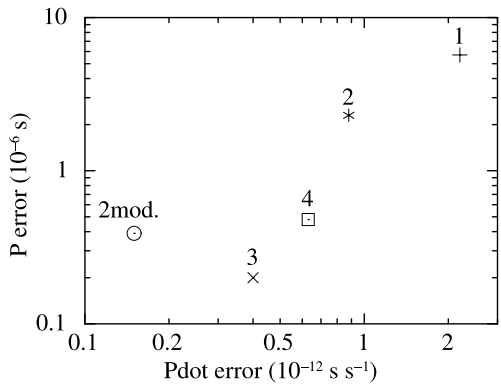}
\end{center}
\caption{Errors of $P$ and $\dot{P}$ obtained by various methods.
The methods are indicated by the numbers in the figure.}
\label{fig:error_plot}
\end{figure}

We estimated the errors ($\Delta P$, $\Delta \dot{P}$) of $P$ and $\dot{P}$ in the folding by several trial methods.
The test case was the MAXI observation of 4U 1626--67 from MJD 55290 to MJD 55350.
The results are tabulated in table \ref{tab:p_pdot_error_tab} and plotted in figure \ref{fig:error_plot}.
We finally trust the values by the Monte-Carlo simulation (Method 4).
Comparing to that, the $\Delta P$ value of Method 2 would be only appropriate if the pulse shape is sinusoidal.
Since the real pulse shapes are sharper than it, it can be a ``loose error'' or a conservative error.
When we consider sharpness of the pulse shape (Method 2 modified), we get closer value, although it is not exactly the same as the true value.
However, on $\Delta \dot{P}$, Method 2 gives the closest value.
Since the validity to use the same $a$ as $\Delta P$ to estimate $\Delta \dot{P}$ is not clear,
 the reason why the sinusoidal case (Method 2) gives good value is not known.
The Monte-Carlo simulation (Method 4) gives 3 times larger $a$ for $\Delta \dot{P}$ than for $\Delta P$, the reason of which is also unclear.
We compare those methods in other span (MJD 56250--56310) as listed in table \ref{tab:p_pdot_error_tab_2}.
The relation of $a$ for $\Delta P$ in Method 4 and for $\Delta \dot{P}$ is almost the same.

\begin{table}
\caption{Errors of $P$ and $\dot{P}$ in MJD 55290--55350.}
\label{tab:p_pdot_error_tab}
\begin{center}
\begin{tabular}{lccc}
\hline
Method & $\Delta P$ & $\Delta \dot{P}$ & $a$ \\
 & (${10}^{-6}$ s) & (${10}^{-12} ~ \rm{ s ~ {s}^{-1} }$) & \\
\hline
1 ``standard'' & 5.7  & 2.2                      & 0.5                                                          \\
2 Leahy        & 2.3  & (0.88)\footnotemark[$*$] & 0.20                                                         \\
2 modified     & 0.39 & (0.15)\footnotemark[$*$] & 0.034                                                        \\
3 pulse fit    & 0.2  & 0.4                      & 0.02\footnotemark[$\dagger$], 0.09\footnotemark[$\ddagger$]  \\
4 MC           & 0.48 & 0.63                     & 0.042\footnotemark[$\dagger$], 0.14\footnotemark[$\ddagger$] \\
\hline
\end{tabular}
\end{center}
\footnotesize
\par\noindent
\footnotemark[$*$] $\Delta \dot{P}$ is not given in \citet{1987A&A...180..275L}.
\par\noindent
\footnotemark[$\dagger$] Calculated from $P$.
\par\noindent
\footnotemark[$\ddagger$] Calculated from $\dot{P}$.
\end{table}

\begin{table}
\caption{Errors of $P$ and $\dot{P}$ in MJD 56250--56310.}
\label{tab:p_pdot_error_tab_2}
\begin{center}
\begin{tabular}{lccc}
\hline
Method & $\Delta P$ & $\Delta \dot{P}$ & $a$ \\
 & (${10}^{-6}$ s) & (${10}^{-12} ~ \rm{ s ~ {s}^{-1} }$) & \\
\hline
1 ``standard''                      & 5.7  & 2.2                      & 0.5                                                     \\
2 Leahy                             & 3.3  & (1.3)\footnotemark[$*$]  & 0.29                                                    \\
2 modified                          & 0.56 & (0.21)\footnotemark[$*$] & 0.049                                                   \\
3 pulse fit\footnotemark[$\dagger$] & --   & --                       & --, --                                                  \\
4 MC                                & 0.90 & 1.07                     & 0.079\footnotemark[$\ddagger$], 0.24\footnotemark[$\S$] \\
\hline
\end{tabular}
\end{center}
\footnotesize
\par\noindent
\footnotemark[$*$] $\Delta \dot{P}$ is not given in \citet{1987A&A...180..275L}.
\par\noindent
\footnotemark[$\dagger$] ${\chi}^{2}$ minimum region could not be determined.
\par\noindent
\footnotemark[$\ddagger$] Calculated from $P$.
\par\noindent
\footnotemark[$\S$] Calculated from $\dot{P}$.
\end{table}

\end{document}